\documentclass[prd,preprint,nofootinbib]{revtex4-1}

\usepackage{amssymb} 
\usepackage{amsmath}
\usepackage{xfrac}
\usepackage{bm}
\usepackage{slashed}
\usepackage{datetime}
\usepackage{braket}
\usepackage[noabbrev]{cleveref}
\usepackage[toc,page]{appendix}

\usepackage{titlesec}
\usepackage{setspace}

\crefrangeformat{equation}{(#3#1#4)-(#5#2#6)}
\numberwithin{equation}{section}

\renewcommand{\L}{\mathcal{L}} 

\newcommand{\hf}{\frac{1}{2}}
\newcommand{\fb}[2]{\left(\frac{#1}{#2}\right)}

\newdateformat{monthyeardate}{%
  \monthname[\THEMONTH] \THEYEAR}

\begin{document}


\hphantom\\
\begin{flushright}
MAN/HEP/2020/14
\\
\monthyeardate{\today}
\end{flushright}

 \title{\Large Quantizing the Eisenhart Lift}

 \author{Kieran Finn\footnote{kieran.finn@manchester.ac.uk}$^*$,
   Sotirios
   Karamitsos\footnote{sotirios.karamitsos@df.unipi.it}$^\dagger$, and
   Apostolos
   Pilaftsis\footnote{apostolos.pilaftsis@manchester.ac.uk}$^*$
   \\
   $\vphantom{invisible}$ }%

 \affiliation{$^*$Department of Physics and Astronomy, University of
   Manchester, Manchester M13 9PL, United Kingdom\\
   \vspace{5mm}
   $^\dagger$Dipartimento di Fisica dell' Universit\'a di Pisa, Italy
 }

\date{\today}
 
\begin{abstract}
  \noindent The classical Eisenhart lift is a method by which the
  dynamics of a classical system subject to a potential can be
  recreated by means of a free system evolving in a higher-dimensional
  curved manifold, known as the lifted manifold.  We extend the
  formulation of the Eisenhart lift to quantum systems, and show
  that the lifted manifold recreates not only the classical effects of
  the potential, but also its quantum mechanical effects. In particular, we
  find that the solutions of the Schr\"odinger equations of the lifted
  system reduce to those of the original system after projecting out
  the new degrees of freedom. In this context, we identify a conserved
  quantum number, which corresponds to the lifted momentum of the
  classical system.  We further apply the Eisenhart lift to Quantum
  Field Theory (QFT). We show that a lifted field space manifold is able to
  recreate both the classical and quantum effects of a scalar field potential. We find that, in the case of QFT, the analogue of the lifted momentum is a
  quantum charge that is conserved not only in time, but also in
  space. The different possible values for this charge label an
  ensemble of Fock spaces that are all disjoint from one
  another. The~relevance of these extended Fock
  spaces to the cosmological constant and gauge hierarchy
  problems is considered.

\end{abstract}

\maketitle 

\section{Introduction}\label{sec:intro}
 
Although it is not often emphasized in the literature, fictitious or
emergent forces underpin a significant part of modern
physics. Fictitious forces arise in Newtonian mechanics in non-inertial frames of
reference such as the surface of the Earth, but they are also the theoretical underpinning of General Relativity
(GR). As demonstrated by Einstein~\cite{Einstein:1915ca}, the effects
of gravity emerge from the curvature of the spacetime manifold. Particles living in anything other than a flat spacetime
can never be in an inertial frame of reference, and, therefore, if we ignore the curvature of
spacetime, gravity must
be introduced as a fictitious force to explain the motion of these particles. In a similar vein, soon after the development of~GR,
Kaluza--Klein theory~\cite{Kaluza:1921tu,Klein:1926tv,
  Overduin:1998pn} was proposed as a way to
explain electromagnetism as a fictitious force. By endowing spacetime
with a compactified fifth dimension, it is possible to show that the additional degrees of freedom in the metric
results in Maxwell's equations.

There is clearly precedent for using the geometrical properties of a
higher-dimensional manifold in order to explain the origin of a
force. As a result, we may naturally wonder whether geometrization is
possible for all such forces. This question was answered by Eisenhart,
who showed that the effects of \emph{any} conservative force on a particle
can be captured by embedding the particle in a higher dimensional
space with \emph{no potential} and an appropriate metric
function~\cite{eisenhart}. This formalism is known as the
\emph{Eisenhart lift}. 

Historically, the Eisenhart lift was the catalyst for the development
of the Bargmann structure framework \cite{Duval:1984cj}, which lifts a
Euclidean manifold with the addition of two extra dimensions such that
it is endowed with a Lorentz metric. In this context, the formalism is
also known as the \emph{Eisenhart--Duval lift}, and has seen
applications in analytical mechanics, particularly in the context of integrable systems~\cite{Cariglia:2013efa,Cariglia:2016oft}; 
see~\cite{Cariglia:2015bla}
for a pedagogical overview. 
 
Recently, we showed that the Eisenhart lift can also be applied to
classical field theories~\cite{Finn:2018cfs}. In this case, the field
theory is ``lifted'', not by the addition of a spatial dimension, but
rather the addition of a new field. This has allowed for a geometric understanding of multifield inflation
\cite{Sasaki:1998ug,Gordon:2000hv,Seery:2005gb}, as well as
the initial conditions of the
Universe~\cite{Finn:2018krt,Finn:2019owf}. 

Thus far, all work done on the Eisenhart lift has been restricted to
the classical level. 
The aim of this paper is, therefore, to consider the quantum effects of the Eisenhart lift.
We will apply the Eisenhart lift to quantum systems in the
contexts of both Quantum Mechanics (QM) and Quantum Field Theory (QFT) and compare the original and lifted systems at the quantum level.

The layout of this paper is as follows: in
Section~\ref{sec:classical_eisenhart}, we provide a brief overview of
the classical Eisenhart lift, including both its original Lagrangian formulation
as well as how it appears in the Hamiltonian formalism. In
Section~\ref{sec:qm_eisenhart}, we apply the Eisenhart lift to QM, demonstrating that solutions to the
lifted Schr\"odinger equation correctly reproduce all the usual
observables of the original system. As an illustrative example\-, we
consider the quantum simple harmonic oscillator and construct its
lifted formulation, as well as the associated lifted Hilbert space. In
Section~\ref{sec:eisenhart_qft}, we extend the Eisenhart
lift to QFTs. We demonstrate that the lifted QFT
contains a conserved quantum charge that labels an ensemble of
disjoint Fock spaces. The value of this charge sets the scale of the
energy quanta of the lifted creation and annihilation operators in the
associated Fock space. We show that for an appropriate choice of the
Fock vacuum, the original space can be embedded in the lifted space in
a similar way to QM. Finally, we discuss how the Eisenhart lift may
help answer the gauge hierarchy and cosmological constant problems in
Section~\ref{sec:hierarchies},  before summarizing our findings and discussing possible further applications in Section~\ref{sec:discussion}.

\section{The Classical Eisenhart Lift}\label{sec:classical_eisenhart}

In this section, we give an overview of the Eisenhart lift as applied
to classical particles. First, we briefly summarize
the Eisenhart lift, as it was originally developed in the Lagrangian
formalism. We then examine how it can be rewritten in the
Hamiltonian formalism, since this will be more convenient for
quantization.

\subsection{Lagrangian Formalism}

To start with, let us  consider a particle of mass $m$ moving in $d$ dimensions. The dynamics of such a particle is described by the Lagrangian
\begin{align}\label{orig}
L  =  \frac{1}{2}  m\sum_{i=1}^d \dot x_i^2  -  V({\bf x}) ,
\end{align}
where 
\begin{align}
{\bf x} = (x_1,\ldots, x_d).
\end{align}
As expected, the evolution of this system is governed by Newton's
second law of motion,
\begin{align}\label{newton2}
m\ddot{x}_i = - \frac{dV({\bf x})}{dx_i}\ .
\end{align}

However, as discussed in the Introduction, it is possible reproduce these equations of motion by means of a \emph{free} particle particle living within a higher-dimensional curved
manifold. To do this, we
introduce a new ``fictitious'' coordinate $y$, and write the
Lagrangian of the lifted system as
\begin{align}
    \label{lifted}
L_{\rm lift} =  \frac{1}{2}  m\sum_{i=1}^d \dot x_i^2\: +\:
  \frac{1}{2}\frac{M^2}{V({\bf x})}\dot y^2 . 
\end{align}
Here we have introduced an arbitrary mass scale $M$ in order to keep dimensions consistent.

Note that in the free limit, $V\to 0$ , the lifted Lagrangian appears to be singular. However, as we will soon see,
$\dot y^2/V$ vanishes in this limit, preventing $L_{\rm lift}$ from
diverging.

The Lagrangian given in \eqref{lifted} has no potential term, and so
the particle is indeed free and will, thus, move along geodesics of this ``lifted''
$(d+1)$-dimensional manifold. As we shall see, these geodesics will recreate the equations of motion~\eqref{newton2}.

By varying \eqref{lifted}, we see that the equations of motion for the lifted system
are
\begin{align}\label{lifted_eom}
m \ddot x_i & = - \frac{M^2}{2}\frac{V_{,i}}{V^2} \dot y^2, 
& 
\frac{d}{dt} \left( \frac{\dot y}{V }\right) &= 0.
\end{align}
The solution to the second equation in \eqref{lifted_eom} is 
\begin{equation}
\label{eq:ydot classical}
\dot y = A\frac{V}{M},
\end{equation}
where $A$ is a dimensionless constant that depends on the initial
conditions.

If we substitute this solution into the first equation
of~\eqref{lifted}, we arrive at
\begin{align}\label{newton_c}
 m \ddot x_i &= -  \frac{ A^2    }{2} V_{,i}.
\end{align}
We see that if the constant $A$ satisfies the Eisenhart condition
\begin{equation}
A^2=2\label{eq:eisenhart condition}
\end{equation}
then~\eqref{newton_c} reduces identically to Newton's second law~\eqref{newton2}.
Thus, once projected down to the original $x$-coordinates, the
geodesics of this lifted manifold reproduce the motion of the particle
described by the original Lagrangian with a potential~\eqref{orig}.

We observe that the equations of motion of the lifted system~\eqref{lifted} are invariant under rescalings of time $t$. This arises because we have the freedom to affinely reparametrize the
geodesics of the lifted manifold. We can always use this freedom to satisfy~\eqref{eq:eisenhart condition} and, hence, recreate the equations of motion of the original system~\eqref{newton2}. If we do not, then the lifted system will still emulate the original system, but will appear to evolve either in ``slow-motion''
($A < \sqrt{2}$) or ``fast-forward'' ($A> \sqrt{2}$).

Finally, we observe that~\eqref{eq:ydot classical} and~\eqref{eq:eisenhart condition} imply that
that $\dot y \to 0$ as $V\to 0$ in such a way that the second term
in~\eqref{lifted} vanishes in this limit, as promised.

\subsection{Geometric Interpretation of the Eisenhart Lift}
 \label{sec:geominterpr}
 
 As mentioned in the Introduction, it is illuminating to view the
 Eisenhart lift from a geometrical point of view. The Lagrangian
 \eqref{orig} describes a particle which lives in Euclidean space $\mathbb{R}^d$ with line element 
 \begin{equation}
 ds^2= \delta_{ij} dx_idx_j.
\end{equation}
 This particle feels an \emph{external} force that is not
 geometrical in origin, but rather due to a potential that permeates
 the manifold.

 On the other hand, the lifted Lagrangian \eqref{lifted} corresponds
 to the motion of a \emph{free} particle on a curved manifold,
 equipped with a metric
\begin{align}\label{eisenhart_metric}
G_{AB}=\begin{pmatrix}
\delta_{ij}&0\\
0&\dfrac{M^2}{mV}
\end{pmatrix}\,
\end{align}
and \emph{generalized coordinates} $q^A$, where $A\in\{i,y\}$ and $q^i
= x^i$, $q^y = y$. The corresponding line element is therefore 
\begin{equation}
ds^2 =  G_{AB} \, dq^A dq^B =   \delta_{ij} \, dx^i dx^j + \frac{M^2}{m V} \ dy^2.
\end{equation}

We note that lowercase indices (corresponding to
the original space) are raised and lowered using the Kronecker delta $\delta_{ij}$. The vertical position of these indices is, therefore, irrelevant. Uppercase indices (corresponding to the lifted system) are, on the other hand, raised and lowered by the metric $G_{AB}$ and its inverse $G^{AB}$. The vertical position of these indices is, therefore, extremely important.

Because the lifted system does not contain a potential the trajectory of the particle will correspond to a geodesic of this manifold. These geodesics can be derived by extremizing the
length of the trajectory in the lifted manifold
\begin{align}
 \Delta s_{\rm lift} =  \int ds =  \int d \lambda \,  \sqrt{G_{AB}
  \frac{dq^A}{d\lambda} \frac{dq^B}{d\lambda} }, 
\end{align}
where $\lambda$ is an affine parameter. 

This leads us to the geodesic equation
\begin{align}
\label{eq:geodesic}
\frac{d^2  q^A}{d\lambda^2} + \Gamma^A_{BC}   \frac{d q^B }{d\lambda}
  \frac{dq^C}{d\lambda} = 0\, ,
\end{align} 
where the Christoffel symbols are given, as usual, by 
\begin{align}
\Gamma^A_{BC}\ =\frac{1}{2} G^{AD}\,\Big(G_{CD,B}+G_{BD,C}-G_{CB,D}\Big)\,.
 \end{align} 
 
Explicitly calculating the Christoffel symbols and substituting them
back in the geodesic equation \eqref{eq:geodesic} returns the lifted
equations of motion \eqref{lifted_eom}, as expected. As we demonstrated in the previous section, these equations recover the original equations of motion when the
fictitious coordinate is projected out.

It is also possible to write the geodesic equation in terms of
covariant derivatives. These are defined with
the help of Christoffel symbols to be
\begin{align}
\label{eq:cov der}
\nabla_C  \dot q^A = \partial_C \dot q^A +  \Gamma^A_{CB}   \dot q^B.
 \end{align}
 
 We may also define the covariant time derivative by multiplying~\eqref{eq:cov der} by $\dot q^C$ and formally using the chain rule.
 This gives us
\begin{align}
D_t \dot q^A \equiv   \ddot q^A +  \Gamma^A_{CB} \dot q^C  \dot q^B.
 \end{align} 
 A similar derivation applies for the covariant derivative $D_\lambda$
 of any parameter~$\lambda$ belonging to the same affine class. We may
 therefore write the geodesic equation as
\begin{align} 
D_\lambda \frac{d   q^A}{d\lambda }  = 0.
\end{align}
Bear in mind that applying a covariant derivative on $q^A$ is
meaningless, as it is not a tensor. The appropriate covariant vector is $\dot q^A$.

\subsection{Hamiltonian Formalism}

The Eisenhart lift was originally formulated in the Lagrangian
formalism. However, since our plan is to study its quantum analogue,
it behooves us to examine it in the Hamiltonian formalism as well.
To this end, we proceed as usual and derive the Hamiltonian $H$ of
a system by means of the standard Legendre transform,
\begin{equation}
H=\ p_A \dot q^A-L,
\end{equation}
where $p_A$ is the conjugate momentum defined as
\begin{equation}\label{pAdef}
p_A\equiv\frac{\partial L}{\partial  \dot q^A} \, .
\end{equation}
Therefore, the Hamiltonian corresponding to \eqref{orig} is
\begin{align}
    \label{originalhamiltonian}
H =\sum_{i=1}^d \frac{p_{i}^2}{2m} + V(x)\, ,
\end{align}
while the one pertinent to the lifted Lagrangian~\eqref{lifted} is
\begin{align}\label{liftedhamiltonian}
H_{\rm lift}  =\ \sum_{i=1}^d \frac{p_{i}^2}{2m}\:+\: \frac{V(x)}{M^2}\,\frac{p_y^2}{2}\ .
\end{align}
Here, $p_i=m\dot{x}_i$ and $p_y = M^2 \dot y /V$. Note that when
$V = 0$, the original and lifted Hamiltonians coincide, which makes
the free limit of Eisenhart lift more apparent in the Hamiltonian
rather than in the Lagrangian formalism.

We now use Hamilton's equations, which may be written in a manifestly
covariant manner as follows:
\begin{align}
    \label{eq:HamiltonEqs}
\dot q^A &= \frac{\partial H}{\partial p_A }\,,
  &
    D_t  p_A&\equiv\ \dot{p}_A -\Gamma^C_{AB}\,\dot{q}^B\,p_C =
              - \frac{\partial U}{\partial q^A}\ ,
\end{align}
where
\begin{align}\label{eq:generalizedpotential}
U = U(q^A) = H|_{p^A = 0}
\end{align}
is the generalized potential of the system.

To derive the second of Hamilton's equation in~\eqref{eq:HamiltonEqs}, we
have employed a covariant expression for the Euler--Lagrange equation
of motion, i.e.
\begin{align}\label{eq:covarianthamiltonderivation}
\frac{d}{dt}\,\bigg( \frac{\partial L}{\partial \dot{q}^A}\bigg)\: -\: \frac{\partial
  L}{\partial q^A}\ =\ D_t \bigg(\frac{\partial L}{\partial
  \dot{q}^A}\bigg)\: +\: \frac{\partial U}{\partial q^A}\ =\ 0\; ,   
\end{align}
as well as taking into account the metric compatibility of the
covariant time-derivative, i.e.~$D_t G_{AB} = 0$. Applied to the Hamiltonian~\eqref{originalhamiltonian} we find
\begin{align}
    \label{eq:HEoMs}
    \dot x_i &= p_i/m\,,
&
\dot p_i &= -m V_{,i},
\end{align}
where the subscript ${,i}$ implies differentiation with respect to $x_i$.

For the lifted Hamiltonian \eqref{liftedhamiltonian}, on the other hand, Hamilton's equations read
\begin{align}
    \label{eq:liftedHEoMs}
\dot x_i &= p_i/m, 
&
 \dot p_i  + M^2 m V_{,i} \frac{p_y^2}{2 } &= 0,
\\
\dot y &= p_y V /M^2,
&
\dot p_y &= 0,
\end{align}
with $i\ne y$. Note that for the lifted system, $U = 0$, since the
dynamics are entirely captured by the kinetic term.

We observe that the
lifted momentum $p_y = A M $ is a constant of motion.  Physically,
this is because there is no effective force in the $y$ direction,
since the Hamiltonian has no explicit~$y$ dependence. As before, if we
choose initial conditions such that the constant $A$ satisfies~\eqref{eq:eisenhart condition},
then the lifted Hamiltonian equations of motion
in~\eqref{eq:liftedHEoMs} reduce completely to the original ones
stated in~\eqref{eq:HEoMs}.

It is worth noting that substituting the equation of motion for the
fictitious coordinate in the lifted Hamiltonian
\eqref{liftedhamiltonian} yields the original Hamiltonian
\eqref{originalhamiltonian}. This is in contra\-distinction to the equivalent procedure for the original and lifted Lagrangians~\eqref{orig} and~\eqref{lifted}.

\section{The Eisenhart Lift in Quantum Mechanics}\label{sec:qm_eisenhart}
Now that we have an understanding of how the Eisenhart lift works in classical mechanics, let us turn our attention to its quantum aspects.

We have been calling the additional dimension in the lifted space
\emph{fictitious}. This term suggests that it is impossible to probe
this dimension by any measurement. Indeed, if we live in a subspace of
a higher-dimensional manifold, the only way that a theory and its
lifted counterpart are physically equivalent is if we cannot observe
motion in the $y$-direction. 

In the classical realm, this condition
must be imposed ``by hand.'' After all, there is no inherent reason
why $y$ and $p_y$ are less ``physical'' than $x_i$ and $p_i$ are.

In QM, however, it is possible to distinguish between
compatible and incompatible observables, i.e.~between observables
whose measurements will or will not affect each other. Therefore, if
we can show that the fictitious position and momentum are compatible
with all non-lifted observables, we may then be assured that it is not
possible to directly probe them.

\subsection{Poisson Brackets and Commutator Algebra}

Before quantizing the lifted theory, we need to examine the Poisson
brackets of the classical theory. 

Consider a general theory of a system whose field space is
parametrised by the canonical coordinates $(q^A, p_A)$, where
$q^A = q^A(t)$ and $p_A = p_A(t)$. The dynamics of the system is
governed by the Hamiltonian
\begin{align}
H({  q,p})\ =\ \frac{1}{2}G^{AB} (  q )\, p_A  p_B\: +\: V({  q}),
\end{align}
where the metric and potential have a general dependence on
${q} = (q_1, q_2, \ldots)$.  Note that by the definition
\eqref{pAdef}, $p_A$ is a proper vector. However~$q^A$ is \emph{not} a covariant vector and should instead be treated as a coordinate.  In
general, the (inverse) field space metric $G^{AB}$ can be explicitly
deduced from the Hamiltonian as follows:
\begin{align}
G^{AB} = \frac{\partial^2 H}{\partial p_A \partial p_B}\ .
\end{align}
Note that this definition of $G^{AB}$ enables one to identify metrics for
Hamiltonians that are not necessarily quadratic in $p_A$.

Let us write down the Poisson brackets for the system 
\begin{align}
    \label{standardpoissondef}
\{f, g \}  =   \frac{ \partial f}{\partial q^A} \frac{\partial
  g}{\partial p_A} -  \frac{\partial f}{\partial p_A} \frac{ \partial
  g}{\partial q^B}\ .
\end{align}

It is important to remark here that the usual Poisson brackets as
defined in \eqref{standardpoissondef} are not covariant. If $f$ and
$g$ are tensors in the field space, their transformation properties
will not be inherited by the Poisson bracket $\{ f, g\}$. This means that after
quantization, the operators and their commutators will also be
non-covariant.

However, this will not affect any physical observables, since these are
independent of parametrisation (covariant or not)~\cite{Kunstatter:1986qa}. Therefore, for
the sake of convenience, we will adopt the standard, chart-dependent
formalism for the rest of the paper even though, in doing so, we will unavoidably
lose manifest covariance. A discussion of
how the Poisson brackets may be covariantized is presented in the Appendix.

We now turn our attention to the classical system with
Hamiltonian~\eqref{originalhamiltonian}. In order to simplify the
discussion, we shall consider the case $d=1$, so that the original
system contains only one coordinate $x_1\equiv x$ and one momentum
$p_1\equiv p_x$. After applying the Eisenhart lift, these are
extended by a new fictitious coordinate $y$ and momentum $p_y$.

We can apply~\eqref{standardpoissondef} to the lifted
system~\eqref{liftedhamiltonian} to find the following Poisson
brackets:
\begin{align}
\label{eq:poisson brackets}
\begin{aligned}
  \{x, x\} &= \{y,y\} =\{x, y\}   = 0\,,
\\
 \{p_x, p_x\} &= \{p_y,p_y\} = \{p_x, p_y\}  = 0\,,
 \\
  \{x, p_x\} &= \{y,p_y\}  =  1\,\\
  \{x, p_y\} &= \{y, p_x\} = 0\,.
  \end{aligned}
\end{align}

In order to quantize the theory, we proceed as usual and promote the
coordinates~$x$, $y$, $p_x$ and $p_y$ to operators~$\hat{x}$, $\hat{y}$, $\hat{p}_x$ and $\hat{p}_y$. Thus, the original
Hamiltonian \eqref{originalhamiltonian} and the lifted
Hamiltonian~\eqref{liftedhamiltonian} become the operator-valued
expressions
\begin{align}
    \label{originalhamiltonianop}
\widehat H &= \frac{\hat p_x^2}{2m} + \widehat V\, ,\\
    \label{hamiltonianliftop}
\widehat H_{\rm lift}  &=  \frac{\hat p_x^2}{2m }  +     \frac{\hat
                         p_y^2}{2 M^2 } \widehat V \, , 
\end{align}
where the hats denote operators and $\widehat V \equiv V(\hat x)$.

Following the canonical quantisation procedure, we promote the Poisson brackets~\eqref{eq:poisson brackets}, yielding
\begin{align}
[\hat  x, \hat p_x ] &= i, &
[\hat y, \hat p_y] &= i,
\end{align}
with all other commutators vanishing.

We notice that any operator
constructed out of~$\hat x$ and~$\hat p_x$ commutes with any operator
constructed out of~$\hat y$ and~$\hat p_y$. This indicates that
fictitious and non-fictitious observables are completely compatible,
and the fictitious subspace of a lifted manifold cannot be probed by
observers living on the ordinary (non-fictitious) subspace.

Finally, we note that we do not run into any ordering issues
with $\widehat V$ and $\hat p_y$, since~${[\hat V, \hat
p_y] = 0}$.  This holds for any potential operator $V(\hat x)$. Thus,
no matter what functional form the potential term assumes, there will 
be no kinetic mixing between $\hat p_x$ and $\hat p_y$.

\subsection{The Lifted Schr\"odinger Equation}

We now turn our attention to the time-independent Schr\"odinger
equation (TISE). For the lifted system described by the Hamiltonian
\eqref{hamiltonianliftop}, this is given by
 \begin{align}\label{liftedschr}
-\frac{\Psi_{,xx}}{2 m} - \frac{V(x) \Psi_{,yy}}{2 M^2} = E \Psi,
\end{align}
where $\Psi = \Psi(x,y)$ is the two-dimensional lifted wavefunction
defined as $\Psi = \braket{x,y|\Psi}$ and the subscripts ${,x}$ and ${,y}$ indicate partial differentiation with respect to $x$ and $y$, respectively.
In writing
down  the TISE~\eqref{liftedschr}, we have used the fact that
 \begin{align}
\bra{x} \hat p_y V(\hat x) \ket{\Psi} = \bra{x} V(\hat x)\,  \hat p_y 
   \ket{\Psi} = -iV(x)\, \Psi_{,y}\; ,
\end{align}
for the state vector $\ket{\Psi}$.  
 
In order to solve \eqref{liftedschr} for the lifted wavefunction, we make the ansatz
 \begin{align}
\Psi(x,y) =  \psi(x)  \chi (y)\,,
\end{align}
giving rise to a set of equations that are fully separable,
\begin{align}
    \label{separated}
\left( \frac{  \psi_{,xx}}{2 m} + E \psi \right)  \frac{
   M^2}{V(x)\psi }  =  -\frac{ \chi_{,yy}}{2 \chi} =  P^2, 
\end{align}
where $P = A M $ is a constant with units of momentum.

We immediately see that $\chi(y)$ is a pure momentum mode, as
expected due to the shift symmetry of $\widehat{H}_{\rm lift}$ in~$y$. 
It can therefore be written as
 \begin{align}\label{Lsol}
\chi (y) = e^{i P y } ,
\end{align}
where we have absorbed an overall constant into the normalisation
for $\psi(x)$. This shows us that the constant $P$ is, in fact, the momentum
in the fictitious direction.

We can substitute this result to~\eqref{separated} and we find
 \begin{align}\label{schrodfin}
-\frac{\psi_{,xx}}{2 m} +  \frac{P^2}{2M^2} V(x)\psi = E\psi.
\end{align}
Interestingly enough, if we choose a state satisfying~\eqref{eq:eisenhart condition} so that $P = \sqrt{2} M$, \eqref{schrodfin} becomes precisely the Schr\"odinger equation for
the original system. In this case, the state $\psi$ will represent an energy
eigenstate of the original Hamiltonian~\eqref{originalhamiltonian}.

The lifted wavefunction $\Psi$ should be normalised so that
 \begin{align}
\int dx dy \, \Psi^*  \Psi  = \int dy \,  \int dx  \, \psi^* \psi  =  1.
\end{align}
However, this equation cannot be satisfied if the fictitious
coordinate is allowed to run over the entire real line.  We therefore choose to
compactify the fictitious direction $y$, restricting it to lie in the
range $y\in[0,\ell)$, for some compactification length $\ell$. This has the advantage of restricting the possibilities for
the fictitious momentum to a set of discrete values
 \begin{align}
 P  =   \frac{2 \pi k}{\ell},
\end{align}
where $k = 0,\, \pm1\,, \pm 2\,\dots$ is an additional quantum number.

We notice that, for the special case $k=0$, the effect of the
potential $V(x)$ in \eqref{schrodfin} vanishes. In~this case, the
solutions for $\psi(x)$ are just waves in the observable~$x$
direction.  Instead, if $V(x)$ is non-zero, we must insist that
$k\neq 0$, so the lowest energy state will be obtained for $k=\pm
1$. Indeed, without loss of generality, we can always choose our
compactification scale $\ell$, such that $k=\pm 1$ corresponds to
$P^2=2M^2$ and so satisfies the Eisenhart condition~\eqref{eq:eisenhart condition}.

In summary, solutions to the lifted TISE~\eqref{schrodfin} are
described by two quantum numbers: (i)~the quantum number $k$ of the
conserved momentum $p_y$, and (ii)~the quantum number~$n$ of a system
with quantized energy eigenstates $E_{k,n}$ that result from a
rescaled potential term, $V_k(x) \equiv k^2 V(x)$.  Consequently, the
states that satisfy the lifted TISE~\eqref{schrodfin} with
non-vanishing potential $V(x)$ can be written as
 \begin{align}
\Psi_{k,n} (x,y)= e^{2\pi i k y / \ell }  \psi_{k,n}  (x)  ,
\end{align}
where 
\begin{equation}
\psi_{k,n}(x)=\psi_n\fb{x}{k}
\end{equation}
is the $n^{\rm th}$ energy eigenstate resulting from the rescaled
potential~$V_k(x)$, and $\psi_n$ is the $n^{\rm th}$ energy eigenstate
of the original system. Notice that, as explained above, one has
$k\neq0$, and so the latter expression is well defined.

\subsection{The Lifted Harmonic Oscillator}

Let us examine closer the spectrum of the harmonic oscillator, which
is described by the lifted Hamiltonian
\begin{align}\label{SHOhamiltonianliftop}
\widehat H_{\rm lift}  &=  \frac{\hat p_x^2}{2m }  +     \frac{\hat x^2  \hat p_y^2}{2 \mu   },
\end{align}
where $\mu  = 2M^2/(m\omega^2 )$.
Solving the lifted TISE~\eqref{liftedschr} for this Hamiltonian, we see that the energy spectrum of the lifted harmonic
oscillator is
\begin{align}
    \label{eq:E_kn}
E_{k,n} = \frac{ 2\pi k\, \omega}{M \ell}  \left(n+\frac{1}{2}\right)\,,
\end{align}
with the restriction of energy positivity: $k > 0$. As discussed above,
the compactification length $\ell$ can be chosen to be $\ell=2\pi/M$,
so that for $k=1$ we have the standard spectrum of the harmonic
oscillator.

Let us now analyse the lifted ladder operators, which will prove instructive when we generalise to QFT. The ladder operator approach is a
particularly intuitive way of deriving the spectrum of a quadratic
theory, and it will be helpful to see how it extends to the lifted
case. We remind the reader that in the standard case, the ladder
operators for the 1D quantum SHO are given by
\begin{align}
     \label{standard_a}
   a \equiv \sqrt{\frac{m\omega}{2}} \left(\hat x +
  \frac{ i \hat p_x }{m\omega} \right),\\
     \label{standard_a_dagger} 
a^\dagger \equiv \sqrt{\frac{m\omega}{2}} \left(\hat x - \frac{ i \hat
   p_x}{m\omega} \right), 
\end{align}
where we have suppressed the hats for $a$ and $a^\dagger$.

A feature of the ladder operators is that the Hamiltonian takes on the
simple quadratic form:
$\widehat H = \omega \left(a^\dagger a + \frac{1}{2}\right)$. By
analogy, we expect for the lifted ladder operators $\mathfrak{a}$ and
$\mathfrak{a} ^\dagger$ to have a similar relationship with
$\widehat{H}_{\text{lift}}$. This consideration prompts us to define
the lifted ladders operators as follows as:
 \begin{align}\label{lifted_a} 
\mathfrak{a} \equiv  \sqrt{\frac{m\omega}{2}} \left( \frac{ \hat x
   \hat p_y}{\sqrt{2} M } + \frac{ i \hat p_x }{m\omega} \right) , 
\\ \label{lifted_a_dagger}
\mathfrak{a} ^\dagger \equiv \sqrt{\frac{m\omega}{2}} \left(   \frac{
   \hat x \hat p_y}{\sqrt{2} M }   - \frac{ i \hat p_x }{m\omega}
   \right)\,, 
\end{align}
for which the lifted Hamiltonian can be written as
 \begin{align} \label{lifted_hamiltonian_alt}
\widehat{H}_\mathrm{lift} = \omega \left (  \mathfrak{a}^\dagger
   \mathfrak{a}+ \frac{\hat p_y}{ 2 \sqrt{2 }  M  }   \right). 
\end{align}

The lifted ladder operators obey the following commutation relation:
 \begin{align} 
[\mathfrak{a},\mathfrak{a}^\dagger]  &=  \frac{\hat
                                       p_y}{\sqrt{2}M},\label{eq:a commutation relation} 
\end{align}
We notice that, when acting on an eigenstate of $\hat{p}_y$, the ladder operators will
assume their usual role with $\mathfrak{a}^\dagger$ and
$\mathfrak{a}$ adding and subtracting a quantum of
energy to the system, respectively.Looking at the lifted Hamiltonian~\eqref{lifted_hamiltonian_alt}, we
observe that the eigenvalue of the fictitious momentum operator~$\hat{p}_y$ sets the ground state energy for the simple harmonic oscillator. In addition, the commutation relation~\eqref{eq:a commutation relation} tells us that this eigenvalue also sets the separation between excited states of the harmonic oscillator.

If we apply these operators to an eigenstate of $\hat{p}_y$ with eigenvalue~$p_y = \sqrt{2} M$, i.e. a $k=1$ state, we will recover the standard relations for the quantum harmonic oscillator. Any other eigenstate will shift the ground state energy, as well as the energies of all excited states, by a fixed amount.

Let us understand what this scaling of energy means, by considering the time dependent Schr\"odinger equation
\begin{equation}
\label{eq:TDSE}
i\hbar \frac{d}{dt}\ket{\Psi(t)}=\hat{H}\ket{\Psi(t)}=\sum_i E_i \alpha_i\ket{\Psi_i(t)}.
\end{equation}
Here, $\ket{\Psi(t)}$ is a general state that has been expanded in the energy eigenbasis
\begin{equation}
\ket{\Psi(t)}=\sum_i\alpha_i\ket{\Psi_i(t)},
\end{equation}
where $\Psi_i$ is an energy eigenstate with eigenvalue $E_i$ and~$\alpha_i$ is a coefficient.

We see from~\eqref{eq:TDSE} that we can compensate for a rescaling of energy by rescaling the time coordinate~$t$. Therefore, just as in the classical case, a state not satisfying~\eqref{eq:eisenhart condition} corresponds to a system evolving in fast forward or slow motion.

Notice that
\begin{equation}
[\widehat{H}_{\mathrm{lift}},\hat{p}_y]=0\,,
\end{equation}
and, hence the eigenvalue $p_y$ is a good quantum number. Therefore, a system prepared in an eigenstate of $\hat{p}_y$ will remain in that state. This is important, because it means that there is no danger of tunnelling to a state with $A\neq\sqrt{2}$ once the time parameter has been rescaled appropriately. Hence, the lifted system will evolve identically to the original system at both the classical and quantum levels.

We are also able to introduce ladder operators for the fictitious $y$
momentum, which raise or lower the quantum number $k$. These are defined
 \begin{align} 
\mathfrak{b} &=   e^{ -2\pi i\hat y/\ell} \frac{  {\hat p}_y
               }{\sqrt{2} M } =  \frac{1}{\sqrt{2}M}  \left(
               \frac{2\pi}{\ell }  + {\hat p}_y    \right)  e^{ - 2\pi
               i\hat y/\ell} , 
\\
\mathfrak{b}^\dagger &=   \frac{ {\hat p}_y }{\sqrt{2}M}  e^{ 2\pi
                       i\hat y/\ell} =    \frac{1}{\sqrt{2}M} e^{
                       2\pi i\hat y/\ell} \,  \left( \frac{2\pi}{\ell
                       }  + {\hat p}_y    \right), 
\end{align}
where we note that $e^{2\pi i \hat y /\ell}$ is an operator-valued
expression.

The remaining commutators are
 \begin{align}
   [\mathfrak{a} ,\mathfrak{b} ^\dagger]=-[\mathfrak{a} ^\dagger ,\mathfrak{b}]
&=   \frac{\pi }{  M^2 \ell }\sqrt{\frac{m\omega}{2}} {\hat x}  {\hat
     p}_y e^{2\pi i \hat y/\ell},\\
   [\mathfrak{a}^\dagger ,\mathfrak{b}^\dagger ]=[\mathfrak{a} ,\mathfrak{b} ]
   &=    \frac{\pi }{  M^2 \ell }
     \sqrt{\frac{m\omega}{2}} {\hat x}  {\hat p}_y e^{-2\pi i \hat
     y/\ell},\\
   [\mathfrak{b} ,\mathfrak{b} ^\dagger] &= \frac{ \pi }{\ell  M^2}
                                           \left( \frac{2\pi }{\ell }
                                           + \hat p_y \right).
 \end{align}
 
 We may now construct the lifted Hilbert space in which the states of
 the lifted harmonic oscillator live. We index states using their
 eigenvalues $k$ and $n$ as
  \begin{align} 
\ket{k,n} = \ket{k} \otimes  \ket{n}.
\end{align}
The lifted ladder operators $\mathfrak{a}^\dagger$, $\mathfrak{a}$ and
$\mathfrak{b}^\dagger$, $\mathfrak{b}$, as defined above, act
as follows on the states:
 \begin{align}
\mathfrak{a}^\dagger \ket{k}\otimes \ket{n} &= k \sqrt{n+1}  \ket{k}\otimes \ket{ n+1} ,
\\
\mathfrak{a} \ket{k}\otimes\ket{k} &=  k \sqrt{n} \ket{k} \otimes \ket{ n-1} ,
\end{align}
and 
 \begin{align} 
\mathfrak{b}^\dagger \ket{k}\otimes\ket{n} &= (k+1) \ket{k+1}\otimes\ket{n} ,
\\
\mathfrak{b} \ket{k}\otimes\ket{n} &= k \ket{k-1}\otimes\ket{n}.
 \end{align}
 With the help of these relations, we may express arbitrary
 eigenstates in terms of the ground state by successively applying their
 creation operators as usual.
 
 We note that the original
 ground state $n=0$ corresponds to a \emph{class} of states
 $\ket{k} \otimes \ket{0}$, leading to an ensemble of Hilbert spaces
 indexed by different vacua. The most general state may be written as
 \begin{align}
     \label{eq:liftedQMspectrum}
   \ket{k,n}\ \equiv\
   \ket{k}\otimes \ket{n}\ =\  \frac{1}{k^n}
   \frac{(\mathfrak{a}^\dagger)^n}{\sqrt{n!}}
   \frac{(\mathfrak{b}^\dagger)^k}{k!} \ket{0}\otimes \ket{0}, 
\end{align}
where the vacua are defined as
 \begin{align} 
\mathfrak{a}\ket{k,0} &= 0,\qquad
\mathfrak{b}\ket{0,n} = 0.
 \end{align}
 
 Note that our definition of Hilbert states
 in~\eqref{eq:liftedQMspectrum} ensures their orthonormality, i.e.
 \begin{align} 
 \braket{k',n'| k,n}  
= \delta_{k'k}\, \delta_{n'n}\, .
\end{align}
This completes our treatment of the lifted harmonic oscillator. In the
next section, we will demonstrate how the Eisenhart lift can be
extended to QFTs.

\section{The Eisenhart Lift in Quantum Field Theory}
\label{sec:eisenhart_qft}

So far, we have discussed classical and quantum aspects of the
Eisenhart lift for particles living in spacetime. However, the
Eisenhart lift can also be applied to field theories. In
this section, we will demonstrate this procedure for both classical and quantum field theories.

\subsection{Classical Field Theory}

Before discussing the quantum aspects of the Eisenhart lift for QFTs, let us briefly review how to
lift a field theory at the classical level~\cite{Finn:2018cfs}.

As an illustrative archetype, consider the following Lagrangian (density)
\begin{align}
  \label{origqft}
\mathcal{L} = \frac{\partial_\mu \phi\, \partial^\mu \phi}{2} - V(\phi),
\end{align}
where all Lorentz indices are contracted with the Minkowski metric
$\eta_{\mu\nu} = {\rm diag}(1,-1,-1,-1)$.  The equation of motion for
this theory is simply the Klein--Gordon
equation
\begin{equation}
\partial^2 \phi + V'(\phi) = 0\,,\label{eq:KG}
\end{equation}
with $\partial^2 \equiv \partial_\mu \partial^\mu$ and
$V'(\phi) \equiv dV(\phi)/d\phi$.

In order to lift this field theory, we must introduce a new,
fictitious, vector field~$B^\mu$.  The lifted Lagrangian in this case
is given by
\begin{align}\label{liftedlagrangianqft}
\mathcal{L}_{\rm lift }= \frac{\partial_\mu \phi  \,  \partial^\mu
  \phi}{2} + \frac{M^4}{V(\phi)}\frac{\partial_\mu  B^\mu \,
  \partial_\nu  B^\nu}{2}, 
\end{align}
where once again the $M^4$ factor is used to keep dimensions
consistent.

The equations of motion for the lifted Lagrangian \eqref{liftedlagrangianqft} are
\begin{align}\label{liftedeomqft}
\partial^2 \phi &=  - \frac{ M^4 }{2}\frac{V'}{V^2} (\partial_\mu B)^2,
&
\partial_\mu \left( \frac{M^4}{V}   \partial_\nu  B^\nu \right) &=   0.
\end{align}
We see that the second equation implies that
\begin{equation}
\label{eq:field space def A}
\frac{M^2\partial_\mu  B^\mu}{V(\phi)} = A,
\end{equation}
where $A$ is some (dimensionless) constant. Substituted back into the
first
equation of~\eqref{liftedeomqft}, this gives us
\begin{align} \label{eomLagrQFT}
\partial^2 \phi   + \frac{A^2 }{2}V' = 0 .
\end{align}
If we choose $A = \sqrt{2}$, we recover the original Klein--Gordon
equation~\eqref{eq:KG}, exactly as we did in QM.
Again, we may always satisfy this condition by rescaling the spacetime
coordinates. However, in the case of field theory, we must
rescale not only time, but space as well. This rescaling leaves the lifted equations of motion~\eqref{liftedeomqft} unchanged.

Let us now examine the above derivation in the Hamiltonian
formalism. We note that the Hamiltonian approach to field theory necessarily breaks
manifest covariance by singling out time. Thus, we should not be surprised that the following derivation is not covariant. Nonetheless, we can be assured that it captures the
same physics.

Our first step is to evaluate the canonical momenta for
the fields through the standard relations,
\begin{align}
    \label{defcanonicalmom}
\pi_\phi = \frac{\partial \mathcal{L}}{\partial \dot \phi},
\qquad
\pi_{ \mu}= \frac{\partial \mathcal{L}}{\partial \dot  B^\mu}.
\end{align}
Normally, we would expect five canonical momenta in total, four of
which correspond to the fictitious fields. However, in the case of the
system described by the Lagrangian in~\eqref{liftedlagrangianqft}, the
three spatial components of $B^\mu$ are auxiliary fields with no
kinetic term and so they have no canonical
momentum. Indeed, substituting~\eqref{liftedlagrangianqft}
into~\eqref{defcanonicalmom}, we find
\begin{align}\label{explicitcanonicalmomenta}
\pi_\phi = \dot \phi,
\qquad
\pi_{  0} = \frac{M^4 }{V(\phi)} \partial_\mu  B^\mu,
\qquad
\pi_{  i} = 0,
\end{align}
with only one non-vanishing fictitious momentum. Nonetheless, in order to reproduce the correct equations of
motion, we must refrain from setting $\pi_i = 0$ too early in our
calculations, but only after we have derived Hamilton's equations.

Our next step is to determine the Hamiltonian (density) of the system
by means of a Legendre transform:
$\mathcal{H} = \pi_A \dot \phi^A- \mathcal{L}$. Using
\eqref{explicitcanonicalmomenta} in order to eliminate all time
derivatives of $\phi$ and $ B^0$, we may write the
Hamiltonian for the lifted system as
\begin{align} 
\label{liftedhamiltoniandensity}
\mathcal{H}_{\rm lift }=  \frac{\pi_\phi^2 + ({\pmb \nabla}\phi)^2}{2}  
+  \frac{V(\phi)}{M^4}\pi_0^2 
+ \pi_i \dot  B^i
-   \pi_0 \partial_i  B^i 
\end{align}
where ${\pmb \nabla}$ denotes the spatial three-derivative. Note that
the term involving $\dot B^i$ cannot be eliminated, since it does not
appear in the definition of any conjugate momenta. Moreover, even though we
know that this term will vanish due to the condition $\pi_i = 0$, we
must not remove~it until after we apply Hamilton's equations.

The equations of motion can be derived from Hamilton's equations,
\begin{align}\label{eq:hamiltonsequationsQFT}
\dot \phi^A &= \frac{\delta {\cal H}}{\delta \pi_A},
&
D_t \pi_A &=   - \frac{\delta {\cal U}}{\delta \phi^A},
\end{align}
which we have written in a manifestly covariant manner similar to \eqref{eq:HamiltonEqs} by now identifying the generalized potential \emph{density} as
\begin{align}
\mathcal{U}(q^A) = \mathcal{H}|_{\pi^A = 0},
\end{align}
which is an expression analogous to \eqref{eq:generalizedpotential}. 
The derivation of the second equation of \eqref{eq:hamiltonsequationsQFT}  is similar to that of \eqref{eq:covarianthamiltonderivation}; namely, from the Euler--Lagrange equations, we can derive
\begin{align} 
\begin{aligned}
\frac{d}{dt}\,\bigg( \frac{\partial {\cal L}}{\partial  \dot{\phi}^A}\bigg)\: &-  {\bf \nabla}_i \,\bigg( \frac{\partial {\cal L}}{\partial  ({\bf \nabla}_i\phi^A)}\bigg)\:   -\: \frac{\partial  {\cal L}}{\partial \phi^A}\ 
 =\ D_t \bigg(\frac{\partial {\cal L} }{\partial   \dot{\phi}^A}\bigg)\: +\: \frac{\delta {\cal U}}{\delta \phi^A}\ =\ 0\;,
\end{aligned}
\end{align}
where we must now take the functional derivative of the generalised potential so as to incorporate the gradient terms. 

Using the equations of motion for the Hamiltonian~\eqref{liftedhamiltoniandensity}, we find
\begin{align}\label{eq1}
\dot \phi  &=\pi_\phi,
&
  \dot \pi_\phi + \frac{\pi_0^2 V'(\phi)  }{2 M^4}   &= \pmb\nabla^2 \phi ,
\\
\label{eq2}
\dot  B^0 &= \frac{V(\phi)}{M^4} \pi_0 - \partial_i  B^i,
&
\dot \pi_0 &=  0,
\\
\label{eq3}
\dot  B^i &= \dot  B^i,
&
\dot \pi_i + \partial_i \pi_0 &=  \dot \pi_i     .
\end{align}
These equations can be rearranged into a more familiar form by
substituting the first equation\- of \eqref{eq1} into the second one,
giving
 \begin{equation} \label{eq:H eom phi 2}
\partial^2\phi=-\frac{1}{2} \pi_0^2 V'(\phi)=-\frac{1}{2}\left[
  \frac{\partial_\mu B^\mu}{V(\phi)}\right]^2 V'(\phi), 
\end{equation}
which is identical to \eqref{eomLagrQFT}. Rearranging the first
equation of \eqref{eq2}, we obtain
\begin{equation}
\pi_0=\frac{M^4\partial_\mu B^\mu}{V(\phi)},
\end{equation}
as expected. Finally, combining the second equation of~\eqref{eq2} 
with the second one of \eqref{eq3} yields
\begin{equation}
\label{eq:pi0 conserved}
\partial_\mu\pi_0=0.
\end{equation}
Thus, we have
\begin{equation}
\pi_0=\frac{M^4\partial_\mu B^\mu}{V(\phi)}=AM^2,
\end{equation}
where $A$ is a constant in agreement with~\eqref{eq:field space def
  A}. Substituting this into~\eqref{eq:H eom phi 2} gives the
equations of motion of the original system, provided $A^2=2$, exactly
as we found before in the Lagrangian formalism.

We have seen that Hamilton's equations also reduce to the generalised
Klein--Gordon equation. However, this derivation gives us more insight
into the constant $A$. We see that, up to a rescaling, it is $\pi_0$:
the conjugate momentum of the fictitious degree of freedom $B^0$.

\subsection{Quantum Field Theory}

Let us now quantise the above lifted theory in the canonical way, by
promoting the fields and conjugate momenta to operators. These
operators satisfy the following canonical (equal-time) commutation
relations:
\begin{align}
\begin{aligned}\label{commute}
[\hat{\phi}(x),\hat{\pi}_\phi(y)]&=i\delta^{(3)}(x-y),\\
[\hat{B}^\mu(x),\hat{\pi}_\nu(y)]&=i\delta^\mu_\nu\delta^{(3)}(x-y),\\
[\hat{\phi}(x),\hat{\pi}_\mu(y)]&=[\hat{B}^\mu(x),\hat{\pi}_\phi(y)]=0,
\end{aligned}
\end{align}  
with all other commutators vanishing.

For illustration, let us consider a free theory with a  Hamiltonian,
\begin{equation}
   \label{eq:Hphi2}
H=\hf\partial_\mu\phi\partial^\mu\phi+\hf m^2\phi^2.
\end{equation}
The  equations of motion for this theory are given by
 \begin{align}
 \label{eq:free field eom}
 \partial^2 \phi    +   m^2 \phi = 0.
\end{align}
Therefore, this QFT can be viewed as a simple harmonic
oscillator at every point in space.

To highlight this interpretation, we perform the Fourier
transform of the above operators
\begin{align}\label{fouriertransf}
 \hat \phi({\bf x}) =  \int \frac{d{\bf k}}{(2\pi)^3} e^{i {\bf
  k}\cdot {\bf x}} \,   \hat \phi_{\bf k}.
 \end{align}
 Here, we work in the Heisenberg picture in which the Fourier
 components have no time dependence. Substituting these into the
 Hamiltonian~\eqref{eq:Hphi2} gives
 \begin{align}
 \label{eq:hamilton qft SHO}
\widehat H = \frac{1}{2}\int \frac{d^3 {\bf k}}{(2\pi)^3} \left[ \dot
   \phi_{\bf k}^2 + ({\bf k}^2 +m^2) \hat \phi_{\bf k}^2 \right]. 
 \end{align}
 
The equations of motion for~\eqref{eq:hamilton qft SHO} give us the dispersion relation
\begin{align} 
\ddot \phi_{\bf k} + ({\bf k}^2 + m^2)  \phi_{\bf k} = 0.
\end{align}
 Thus, we see that each Fourier mode oscillates
 as a simple harmonic oscillator with unit mass and frequency
 $\omega_{\bf k} = {\bf k}^2 + m^2$.

Proceeding as in QM, we can define
creation and annihilation operators
\begin{align} 
a_{\bf k}&= \int  d^3 {\bf x} \,  e^{i {\bf k} \cdot {\bf x}} \left[ \sqrt{ \frac{ \omega_{\bf k}}{2}} \hat \phi({ \bf x})  + \frac{i}{\sqrt{2\omega_{\bf k}}}\hat \pi  ({\bf x}) \right] ,
\\
a_{\bf k}^\dagger &= \int  d^3 {\bf x} \,  e^{i {\bf k} \cdot {\bf x}} \left[ \sqrt{ \frac{ \omega_{\bf k}}{2}} \hat \phi({ \bf x})  - \frac{i}{\sqrt{2\omega_{\bf k}}} \hat\pi  ({\bf x}) \right],
\end{align}
where we suppress hats for $a_{\bf k}$ and $a_{\bf k}^\dagger$.
The canonical commutation relations for $\hat{\phi}$ and $\hat{\pi}$ imply that the creation and annihilation
operators obey the following commutation relations:
\begin{align}
    \label{eq:akaq}
[a_{\bf k}, a_{\bf q}] &= [a_{\bf k}^\dagger, a_{\bf q}^\dagger] = 0,\\
    \label{eq:akaqdagger}
[a_{\bf k}, a^\dagger_{\bf q}] &= (2\pi)^3 \delta^{(3)} ({\bf k}- \bf {q}).
 \end{align}
 
 With $a_{\bf k}$ and $a_{\bf k}^\dagger$ thus defined, we can rewrite the Hamiltonian in the form
\begin{align}
\widehat H = \int \frac{d^3 {\bf k}}{(2\pi)^3} \frac{\omega_{\bf
    k}}{2} \big[a_{\bf k}a_{\bf k}^\dagger + a_{\bf k}^\dagger a_{\bf
    k} \big]. 
\end{align}
Note that the usual divergence term $\delta^{(3)}(0)$ resulting from
the commutator~\eqref{eq:akaqdagger} can be taken care of with the
standard normal ordering.

We are now in the position to apply the Eisenhart lift to this free
theory. We see from~\eqref{liftedhamiltoniandensity} that,
classically, this gives us the lifted Hamiltonian
\begin{equation}
\label{eq:lifted free hamiltonian}
\mathcal{H}_{\rm lift }=  \frac{\pi_\phi^2 + ({\pmb \nabla}\phi)^2}{2}  
+  \frac{m^2\phi^2}{2M^4}\pi_0^2 
+ \pi_i \dot  B^i
-   \pi_0 \partial_i  B^i.
\end{equation}

Let us now quantise this theory using the framework of canonical
quantisation. Taking our inspiration from the lifted quantum harmonic
oscillator in Section~\ref{sec:qm_eisenhart}, we define the lifted
creation and annihilation operators
\begin{align} 
\mathfrak{a}_{\bf k}&= \int  d^3 {\bf x} \,  e^{i {\bf k} \cdot {\bf
                      x}} \left[ \frac { \hat \pi_0({\bf
                      x})}{\sqrt{2}M^2} \sqrt{ \frac{ \omega_{\bf
                      k}}{2}} \hat \phi({ \bf x})  +
                      \frac{i}{\sqrt{2\omega_{\bf k}}}\hat \pi_\phi
                      ({\bf x}) \right] , 
\\
\mathfrak{a}_{\bf k}^\dagger &= \int  d^3 {\bf x} \,  e^{i {\bf k}
                               \cdot {\bf x}} \left[ \frac {\hat
                               \pi_0({\bf x})}{\sqrt{2} M^2} \sqrt{
                               \frac{ \omega_{\bf k}}{2}} \hat \phi({
                               \bf x})  - \frac{i}{\sqrt{2\omega_{\bf
                               k}}}\hat  \pi_\phi  ({\bf x}) \right]. 
 \end{align} 
This allows us to rewrite the lifted Hamiltonian in the form
\begin{align}
\widehat H_{\rm lift} = \int \frac{d^3 {\bf k}}{(2\pi)^3}
    \frac{\omega_{\bf k}}{2} \big[\mathfrak{a}_{\bf
    k}\mathfrak{a}_{\bf k}^\dagger + \mathfrak{a}_{\bf k}^\dagger
    \mathfrak{a}_{\bf k} \big]. 
\end{align}
Observe that the form for $\widehat H_{\rm lift}$ is analogous to
\eqref{lifted_hamiltonian_alt} in QM. We note that the
last two terms in \eqref{eq:lifted free hamiltonian} do not appear
because they vanish once integrated over.

Upon normal ordering of the Hamiltonian where all creation operators
occur before annihilation operators, we have
\begin{align} 
:\!  \widehat H_{\rm lift} \! : = \int \frac{d^3 {\bf k}}{(2\pi)^3}
     \omega_{\bf k}    \mathfrak{a}_{\bf k}^\dagger \mathfrak{a}_{\bf
     k} \;. 
\end{align}
The following commutation relations may then be derived:
  \begin{align}
[ \mathfrak{a}_{\bf k},\mathfrak{a}^\dagger_{\bf q}] &= \frac{\hat
                                                       \pi_{0 {\bf k}}
                                                       }{\sqrt{2} M^2}
                                                       (2\pi)^3
                                                       \delta^{(3)}({\bf
                                                       k-q}) 
\\
[ :\!  \widehat H_{\rm lift} \!  :,\mathfrak{a}^\dagger_{\bf k}] &=
                                                                   \frac{\hat
                                                                   \pi_{0
                                                                   {\bf
                                                                   k}}
                                                                   }{\sqrt{2}M^2}
                                                                   \omega_{\bf
                                                                   k}
                                                                   \mathfrak{a}^\dagger_{\bf
                                                                   k}, 
\\ 
[:\! \widehat H_{\rm lift} \!   :,\mathfrak{a} _{\bf k}] &= -
                                                           \frac{\hat
                                                           \pi_{ 0
                                                           {\bf k}}
                                                           }{\sqrt{2}M^2}
                                                           \omega_{\bf
                                                           k}
                                                           \mathfrak{a}
                                                           _{\bf k}. 
 \end{align}
 
We note that if we are in an eigenstate of $\hat{\pi}_0$ with the same
eigenvalue $\pi_0=\sqrt{2}M^2$ for \emph{all} wavenumbers, these
commutation relations reduce to their standard expressions~\eqref{eq:akaq} and~\eqref{eq:akaqdagger}.

To fully understand the states of the lifted system, we calculate the four-momentum operator $\widehat{P}^\mu$. To this end, we
first derive the classical four-momentum $P^\mu$ from the canonical
energy-momentum tensor,
 \begin{align}
T^{\mu\nu} = \frac{\partial \mathcal{L}}{\partial (\partial_\mu\phi)^A} \partial^\nu \phi^A- \eta^{\mu\nu}\mathcal{L}.
\end{align}
Spacetime translational invariance of the original Lagrangian
implies that the four-momentum, 
 \begin{align}
P^\mu = \int\!d^3 x\;T^{0\mu}\,,
\end{align}
is a conserved Noether's charge. 

For the lifted system~\eqref{eq:lifted free hamiltonian},
the energy-momentum tensor is found to be
 \begin{align}
T^{\mu\nu} &= 
  \partial^\mu \phi  \, \partial^\nu \phi 
+ \frac{1}{2} \frac{2M^4}{m^2\phi^2} (\partial_\rho  B^\rho) (
   \partial^\mu  B^\nu  +   \partial^\nu  B^\mu ) 
- \eta^{\mu\nu}\left[  \frac{(\partial \phi)^2 }{2} +
   \frac{2M^4}{m^2\phi^2}\frac{(\partial_\rho  B^\rho)^2
   }{2}\right]. 
\end{align}
In particular, the elements that contribute to the four momentum are
 \begin{align}
T^{00} &= 
 \frac{\dot\phi^2 + ({\pmb \nabla}\phi)^2}{2}
+   \frac{2M^4}{m^2\phi^2} (\partial_\rho  B^\rho)^2.
\\
T^{0i} &= 
 \dot\phi \, \partial^i \phi 
+ \frac{1}{2} \frac{2M^4}{m^2\phi^2} (\partial_\rho  B^\rho) ( \dot B^i  +   \partial^i  B^0 ).
\end{align}
After promoting the fields to operators, these become 
 \begin{align}
\widehat T^{00} &= 
 \frac{\hat \pi_\phi^2 + ({\pmb \nabla}\hat\phi)^2}{2}
+  \frac{m^2\hat \phi^2}{2M^4} \hat \pi_0^2.
\\
\widehat T^{0i} &= 
 \hat \pi_\phi \, \partial^i\hat \phi 
+ \frac{1}{2} \hat  \pi_0 ( \hat \pi_i +   \partial^i \widehat  B^0 ).
\end{align}
Integrating these expressions over the spatial volume,
we obtain the Noether-charge operators${}$:
 \begin{align}
\widehat P^0 &=   {\widehat H} = \int \!d^3 x\; \mathcal{\widehat H},
\\
\widehat P^i &= \int\!d^3 x\; \left[ 
 \hat \pi_\phi \, \partial^i \phi 
+ \frac{1}{2} \hat  \pi_0 ( \hat \pi^i +   \partial^i  \widehat B^0 ) \right].
\end{align}
By employing the commutation relation
\begin{equation}
  [\hat\pi_0,\partial^i\hat{B}^0]=\partial^i[\hat\pi_0,\hat{B}^0] -
  [\partial^i\hat\pi_0,\hat{B}^0]=0\, ,
\end{equation}
which can be deduced thanks to~\eqref{eq:pi0 conserved},
we find that 
 \begin{align}
 \label{Pmu}
[\widehat H, \hat \pi_0]  &= 0,
\\
\label{eq:Pi com}[\widehat P^i, \hat \pi_0]  &= 0.
\end{align}

The vanishing of the commutator in~\eqref{Pmu} implies
that the eigenstates of the operator~$\hat{\pi}_0$ are compatible with
those of the Hamiltonian, and so its eigenvalue~$\pi_0$ represents
good quantum number, which is conserved in time. 

In addition, the vanishing of the commutator~\eqref{eq:Pi com} implies
that this quantum number is also the same throughout all space. Thus,
the eigenvalue $\pi_0$, represents a global quantum number that is the
same everywhere in the Universe.

This conserved charge can be interpreted in two ways; either as the
definition of  our system of (energy) units if we choose to assume a
fixed particle mass, or as the mass of the particle if we assume a
fixed system of units. In the latter case, jumping to an excited state
of $\hat{\pi}_0$ would be interpreted as a change in the mass of the
$\phi$ particle. However~\eqref{Pmu} and~\eqref{eq:Pi com} ensure that
once $\pi_0$ is set by some initial conditions, its value will remain
constant throughout space and time. Therefore, the spectrum of the
theory will not suddenly change as the system evolves.

We can now examine the Fock space of the lifted theory. In the
standard case, the Fock space is the direct sum of tensor products of
copies of the same Hilbert space. A general state in the original Fock
space can be written in slightly expanded notation as 
\begin{align}
\ket{\psi} =  \sum_{\pi_0} \ \ket{(n_1)_{ {\bf k}_1}, (n_2)_{{\bf k}_2}, \ldots},
 \end{align}   
denoting $n_i$ quanta of $\phi$ all with momentum ${\bf k}_i$.

Once the system is lifted, however, the Fock space of the theory is
now extended to an ensemble of Fock spaces, indexed by the value of
$\pi_0$. This is similar to the case we saw in QM. Explicitly, an
eigenstate of the system can be written 
\begin{align}
\ket{\Psi}_{\pi_0} =  \ket{\pi_0 } \otimes  \ \ket{(n_1)_{ {\bf k}_1}, (n_2)_{{\bf k}_2}, \ldots}  ,
 \end{align}
where
 \begin{align}
 \ket{\pi_0 } \otimes  \ \ket{(n_1)_{ {\bf k}_1}, (n_2)_{{\bf k}_2}, \ldots}    &=
  \frac{\left(\mathfrak{a}^\dagger_{\mathbf{k}_1}\right)^{n_{\mathbf{k}_1}}}{\sqrt{n_{\mathbf{k}_1}!}}
  \frac{\left(\mathfrak{a}^\dagger_{\mathbf{k}_2}\right)^{n_{\mathbf{k}_2}}}{\sqrt{n_{\mathbf{k}_2}!}}\ldots 
\times
 \ket{\pi_0 } \otimes \ket{0_{ {\bf k}_1}, 0_{{\bf k}_2}, \ldots},
 \end{align}
 and the vacua are defined so that
  \begin{align}
a_{\bf k} \ket{\pi_0 } \otimes \ket{0_{ {\bf k}_1}, 0_{{\bf k}_2}, \ldots}=0,\;\;\;\forall \bf k.
 \end{align}
These vacua are orthogonal, since the ladder operators commute
with~$\hat{\pi}_0$. Thus, the eigenvalue $\pi_0$ labels an ensemble of
independent Fock spaces that are completely separate from one
another. Note that 
\begin{equation}
\hat{\pi}_0\ket{\Psi}_{\pi_0}=\pi_0\ket{\Psi}_{\pi_0}
\end{equation}
and so $\ket{\Psi}_{\pi_0}$ is an eigenstate of the operator $\hat{\pi}_0$.

The most general state can be written as a superposition of states in
different Fock spaces: 
  \begin{align}
  \label{eq:fock superpos}
\ket{\Psi} = \sum_{\pi_0} \alpha_{\pi_0} \ket{\Psi}_{\pi_0}.
 \end{align}
Thanks to the orthogonality of the Fock spaces, any observer will only
have access to a single slice of this state, corresponding to a single
value of $\pi_0$. Different observers will measure different values of
$\pi_0$ and, hence, different values for the mass of $\phi$, but these
observers will never be able to communicate with one another. 
 
Therefore, we can interpret the state~\eqref{eq:fock superpos} as a
superposition of different ``universes'' with different masses for the
particles, all of which evolve completely independently. We happen to
live within one such universe, and will forever stay within it, since
the states are orthogonal. 
 
\section{Dynamical Generation of Hierarchies via the Eisenhart Lift}\label{sec:hierarchies}

The gauge hierarchy problem \cite{Gildener:1976ai,Weinberg:1978ym} and
the cosmological constant problem \cite{Weinberg:1988cp} are two of
the most puzzling aspects of modern particle physics and indicate an
extreme level of \emph{fine-tuning} in the Standard Model. In absence
of a mechanism that can explain how they arise, a very delicate
balance must be struck such that radiative corrections cancel out to
match measurements. This is compounded by the fact that such
cancellations are sensitive to short-scale physics that remain
unprobed. These issues do not necessarily indicate an underlying
problem with our current theories, but rather are signs pointing to
the existence of a more complete theory. 

There have been many attempts to resolve the gauge hierarchy and the
cosmological constant problems. The most prominent of these involve
theories beyond the Standard Model such as
Supersymmetry~\cite{Dimopoulos:1981zb}. However, recent observational
bounds on supersymmetric parameters lead to a similar level of
fine-tuning, at least as far as the cosmological constant problem is
concerned~\cite{Draper:2011aa}. 

A possible way to evade the cosmological constant problem is by
dynamical generation of the cosmological constant, such as via
quintessence models, in which a scalar field takes on the role of dark
energy \cite{Wetterich:1987fm,Ratra:1987rm, Caldwell:1997ii}. In these models, the
density of the field eventually dominates the expansion rate of the
Universe by rolling down a potential \cite{Zlatev:1998tr}. Even then,
such models cannot fully evade fine-tuning: careful parameter matching
is required in order to generate the low scale of the vacuum energy
observed today \cite{LopesFranca:2002ek}. 

Another method to resolve fine-tuning issues comes from the anthropic
principle. Fine-tuning was originally identified by Dicke in relation
to the age of the Universe~\cite{dicke1961}, who noted that living
observers must be necessarily observe the age of universe to within a
window of opportunity that is neither too young nor too old. This
argument was codified in the so-called \emph{anthropic principle}~\cite{Carter:1974zz} 
and used as a heuristic to argue that  physical
and cosmological quantities are necessarily biased towards values that
can support intelligent observers. This provides an explanation for
the apparent ``cosmic coincidences'' of our universe (such as the
present gauge hierarchy and the small value of the cosmological
constant); if these coincidences did not exist, we would not be here
to observe them.  

While the anthropic principle has often been criticized as a truism,
it becomes much more useful when coupled with a landscape of theories,
some of which can support observers and some of which cannot, and from
which a particular universe can be anthropically selected. An example
of this is the \emph{multiverse}~\cite{Kachru:2003aw,Weinberg:2005fh}, which
provides such a landscape. Even then, anthropic arguments remain
controversial, since they offer no new physical
insights~\cite{Smolin:2004yv} and because assigning statistical
weights to different universes can be an ambiguous
procedure~\cite{Starkman:2006at,Gibbons:2006pa}.   
 
The Eisenhart lift may provide an alternative solution to both the
gauge hierarchy and the cosmological constant problems without
reference to short-scale physics. The gauge parameters, instead of
being imposed \emph{ad hoc}, can now be thought of as part of the
geometry of the field space of our theory. By absorbing the offending
quadratic divergence terms into the metric, it becomes possible to
tame them by choosing an appropriate subspace. This completely
eliminates this kind of fine-tuning since, as discussed in the
previous subsection, the extended Fock space ensemble incorporates all
possible particle masses.

The extended Fock space can incorporate all possible spectra of
different particles simply by adding more fictitious fields. As a
result, the observed hierarchy depends wholly on which slice we live
on. This solves the issue of finely tuned parameters in dynamical
models. 

We can also describe a truly constant vacuum energy using the Eisenhart lift, resulting in a Lagrangian of the form
\begin{equation}
\L=\L_{\rm SM}+\frac{1}{2M_P^2} \nabla_\mu B^\mu\nabla_\nu B^\nu,
\end{equation}
where~$\L_{\rm SM}$ is the Lagrangian of the Standard Model (minus the
cosmological constant term),~$M_P$ is the Planck mass and~$B^\mu$ is
the fictitious field. The cosmological constant problem would then be
solved if we lived on a slice with $\pi_0\sim10^{-60}M_P$. 

Alternatively, we may live in a slice where $\pi_0=0$, leading to an
exactly vanishing cosmological constant. The small observed value
could then be generated by suppressed anomalous tunneling effects
between slices.

The Eisenhart lift resolves fine-tuning issues in a way that combines
both anthropic considerations and the geometry of the field space. In
fact, the Fock space ensemble is closely analogous to the multiverse,
in that they both allow for the realization of a landscape of
theories. The crucial difference is that the multiverse spans the
entirety of spacetime, while the ensemble spans the entirety of
possible QFTs. As with any anthropic argument, we must necessarily
contend with the difficulties of assigning an indifferent prior to an
infinite landscape. However, even in the face of this issue, the
Eisenhart lift provides a novel avenue for dealing with problems of
fine-tuning.

\section{Discussion}\label{sec:discussion}

In this work, we have studied the Eisenhart lift and its applications
to quantum theory. By reformulating the lift in the Hamiltonian
formalism, we have been able to apply it to QM and QFT. We have shown
that the introduction of a additional, ``fictitious'' degrees of
freedom, which replicate the effects of a conservative force in a
purely geometric manner in the classical case, also replicate the
observables of the corresponding quantum theory. This applies both
when quantizing particles in QM and when quantizing fields in QFT. 

In order for the Eisenhart lift to fully reproduce the equations of
motion of the original theory, we must impose a particular value for
the fictitious momentum. This degree of freedom stems from the simple
fact that the lifted manifold is higher dimensional. However, it does
not spoil the correspondence between a system and its lifted
counterpart. This is because geodesics in the lifted manifold are
affinely reparametrizable, and so different choices of the fictitious
momentum still lead to the same trajectories in the original manifold,
with the only difference being a different scaling of time or, more
generally, choice of units. This is reflected in the lifted quantum
mechanical system~\eqref{lifted}, where different fictitious momenta
correspond to different masses for the particle. Therefore, an
Eisenhart-lifted system will always correspond to an \emph{equivalence
  class} of systems.  
 
In QM, the additional degrees of freedom introduced by the Eisenhart
lift is reflected in the choice of the vacuum. Quantizing the lifted
theory results in the same results as the original theory (with the
same probability distributions and spectrum) only if the vacuum is
appropriately selected. However, different eigenstates of the
fictitious momentum evolve separately, which means that once again,
the lifted manifold gives rise to an equivalence class of Hilbert
spaces that differ in their zero-point energy. Similarly, in QFT, the
lifted manifold corresponds to an equivalence class of Fock spaces,
with different values for the mass of the original particle. States
that belong in different subspaces of the equivalence class evolve
separately. Therefore, observers in the real world effectively live in
one of the ``slices'' of the lifted Hilbert space. 
 
We have focused on the simple harmonic oscillator in QM and free
particles in QFT, since the ladder operator formalism makes their
study somewhat easier. However, the Eisenhart lift is not restricted
to particular potentials. In the spirit of perturbation theory, we may
use the free theory as a starting point and slightly deform it by a
small potential term to find the relevant observables. Iterating this
procedure will allow us to approximate any potential we like. Using
this technique, it is in principle possible to derive expressions for
the ladder operators of any
potential~\cite{Jalali:2013,Cardoso:2017qmj}. 

With the development of the quantum formulation of the Eisenhart lift,
the scope of possible applications is considerably expanded. We have
seen how the dynamic generation of hierarchies offered by the
Eisenhart lift may provide a novel way to tackle the gauge hierarchy
and cosmological constant problems. It would then be interesting to
examine how symmetries and gauge fields translate to a lifted theory,
paving the way to a complete geometrical interpretation of the
Standard Model.  

Finally, our formulation has focused on scalar fields, but we expect
the Eisenhart lift to be applicable to tensors~\cite{Finn:2019aip} as
well as spinors~\cite{Finn:2020nvn}, possibly paving the way to study
not only the Standard Model through the lens of geometry, but also
providing a fundamentally different insight to our attempts to
quantize gravity.

\begin{acknowledgements}
The work of KF was supported in part by the University of Manchester
through the President's Doctoral Scholar Award. The work of SK was
supported by the ERC grant 669668 NEO-NAT. The work of AP is supported
by the Lancaster--Manchester--Sheffield Consortium for Fundamental
Physics under STFC research grant ST/L000520/1. 
\end{acknowledgements}

\appendix

\section{Covariant Poisson brackets}

In order to arrive at a covariant definition Poisson brackets in
curved space we try promoting the ordinary derivatives with respect to
the coordinates to covariant derivatives:. This leads us to define 
\begin{align}\label{poissondef}
\{f, g \}_G =   \nabla_A f \frac{\partial g}{\partial p_A} -
  \frac{\partial f}{\partial p_A} \nabla_A g. 
\end{align}
We have used the subscript $G$ to remind us that this definition of
the Poisson bracket depends on the metric $G_{AB}$. We can easily see
that this definition respects the tensor properties of $f$ and $g$. In
addition, for flat space the covariant derivatives reduce to partial
derivatives and we recover the usual result. 
 
However, defining the covariant Poisson brackets does not fully
resolve the issue of non-covariance. While we may be tempted to find
the covariant commutator $\{q^A,p_B\}_G$, we must remember that $q^A$
is not a field space vector, but rather a field space coordinate
belonging to a particular chart. Therefore, $\{q^A,p_B\}_G$ is a
non-covariant quantity despite the use of~\eqref{poissondef}. In fact,
as noted in Section~2.2, $\nabla_B q^A$ is a meaningless
expression. This means that in order to write down Poisson brackets in
a covariant manner, we must use a vector in place of $q^A$ .  

It is possible to promote $q^A$ to a vector in a manner similar to
that used in the (famously covariant) Vilkovisky--De Witt
formalism~\cite{DeWitt:1967ub,Vilkovisky:1984st,Vilkovisky:1984un}. We
can define the \emph{tangent vector} $\sigma^A(q_*,q)$ by considering
an arbitrary base point $q_*$. 

 The tangent vector $\sigma^A(q_*,q)$ is a scalar with respect to its first argument and a vector with respect to its second argument, and can be expanded as
\begin{align}
\nonumber
 \sigma^A(q_*,q) =   &- \big(q^A_* - q^A \big) \! - \! \frac{1}{2} \Gamma^A_{BC}(q_*)\,\big(q^B_*  \! - \! q^B\big)\big(q_*^C \! - \! q^C\big)
 \\
    \label{sigmaexpansion}
 &+\: \mathcal{O}\big[(q_*-q)^3\big].
\end{align}

We then see that $\{ \sigma^A(q_*,q), p_B\}_G$ is a tensor due to the fact that~$q_*$ does not depend on~$q$ or~$p$. Taking into account the curvature of the space, we may write the following tensor expressions for the Poisson brackets, suppressing the arguments for $\sigma^A(q_*,q)$:
 \begin{align}
 \label{pois1}
\{\sigma^A, \sigma^B  \}_G &= \{p_A, p_B\}_G  = 0,
\\
\label{pois2}
\{\sigma^A , p_B\}_G &= \delta^A_B \;- \; \frac{1}{3} R^A_{\ CBD} \; \sigma^C (q_*,q )  \; \sigma^D(q_*,q) \; +\; \ldots.
\end{align}
We observe that the commutator in \eqref{pois2} is not $\delta^A_B$ as
we expect, but is modified by curvature terms induced by $G_{AB}$. 

Once we know the Poisson structure of the theory, we can canonically quantize by using
\begin{align}\label{eq:quantmap} 
\{f, g\}_G \rightarrow -i [\hat f, \hat g]_G,
\end{align}
where $[\hat f, \hat g]_G$ is defined such that it reduces to the
standard commutator $[\hat f,\hat g] = \hat f \hat g - \hat g \hat f$
for a flat space.

We also need to find covariant representations for the operators
$\hat{p}_A$ and $\hat{\sigma}^A$. We suspect that the momentum
operator will be represented by $\hat p_A = -i\nabla_A$, but the
representation of~$\sigma^A$ is not as readily apparent. We will not
address here the covariant quantization of the quantum operator
e.g. the mapping $p_A \to \hat p_A = -i\nabla_B$ or the role of
$\sigma^A$, but shall instead leave it for future work.

\end{document}